\documentclass{ws-fnl}
\usepackage{cite}
\usepackage{url}
\begin{document}

\def\bib{B\kern-.05em{I}\kern-.025em{B}\kern-.08em}
\def\btex{B\kern-.05em{I}\kern-.025em{B}\kern-.08em\TeX}
%

\catchline{}{}{}{}{}
\title{
Revisiting the Boltzmann derivation 
of the Stefan law
}
\author{Lino Reggiani}
\address
{Dipartimento di Matematica e
Fisica, "Ennio De Giorgi" \\ Universit\`a del Salento, via Monteroni
73100 Lecce, Italy.
}

\author{Eleonora Alfinito}
\address
{Dipartimento di Matematica e
Fisica, "Ennio De Giorgi" \\ Universit\`a del Salento, via Monteroni
73100 Lecce, Italy.
}
\date{\today}
\maketitle
\begin{history}
\received{(received date)}
\revised{(revised date)}
\end{history}
\begin{abstract}
The Stefan-Boltzmann (SB) law relates the  radiant emittance
 of an ideal black-body cavity at thermal equilibrium
to the fourth power of the absolute temperature  $T$ as
$q=\sigma T^4$, with $\sigma = 5.67 \times 10^{-8} \ W m^{-2} K^{-4}$ the SB constant, firstly estimated by Stefan to within $11$  per cent  of the present theoretical value.
The law is an important achievement  
of modern physics since, following Planck (1901), its microscopic derivation implies the quantization of the energy related to the electromagnetic field spectrum. 
Somewhat astonishing, Boltzmann  presented his derivation in 1878 making use only of electrodynamic and thermodynamic classical concepts, apparently without introducing any quantum hypothesis (here called first Boltzmann paradox).
By contrast, the Boltzmann derivation implies two assumptions not justified within a classical approach, namely: (i)  the zero value of the chemical potential and, (ii) the internal energy of the black body with a finite value and dependent from both temperature and volume.
By using Planck (1901) quantization of the radiation field  in terms of a gas of photons, the SB law received a microscopic interpretation free from the above assumptions that also  provides the value of the SB constant on the basis of a set of universal constants including the quantum action constant $h$. 
However, the successive consideration by Planck (1912) concerning the zero-point energy contribution was found to be responsible  of another divergence of the internal energy for the single photon mode at high frequencies.  This divergence is of pure quantum origin and is responsible for a vacuum-catastrophe, to keep the analogy with the well-known ultraviolet catastrophe of the classical black-body radiation spectrum, given by the Rayleigh-Jeans law in 1900. As a consequence, from a rigorous quantum-mechanical derivation we would expect the divergence of the SB law (here called second Boltzmann paradox).
Here, both the Boltzmann paradoxes are revised by accounting for both the quantum-relativistic photon gas properties, and the Casimir force.
\end{abstract}
\section{Introduction}
The Stefan-Boltzmann (SB) law \cite{stefan,boltzmann} refers to the radiant emittance  of an ideal black-body cavity at a given temperature $T$ under thermal equilibrium and thermodynamic limit conditions 
\cite{crepeau09}.
It takes the form:
\begin{equation}
q=\sigma T^4
\label{eq1}
\end{equation}
with $q$ the radiant emittance given in $Wm^{-2}$, $T$ the absolute 
temperature in Kelvin and 
\begin{equation}
\sigma = 5.67 \times 10^{-8} \ W m^{-2} K^{-4}
\label{eq2}
\end{equation}
the SB constant, here given with an accuracy of three digits, and firstly estimated by Stefan in 1874 on the basis of available experimental data with an accuracy of $11$ per cent.
\par
The SB constant is actually expressed in terms of universal constants as:
\begin{equation}
\sigma=\frac{2\pi^5 K_B^4}{15 h^3 c^2}
\label{eq3}
\end{equation}
with $K_B$ the Boltzmann constant, $h$ the Planck constant and $c$  the light speed in vacuum.
Accordingly, the SB law is an important achievement of today modern physics 
(i.e. quantum mechanics and relativity), at the time of its formulation unknown by both Stefan and Boltzmann. 
Yet, we notice the paradox, that apparently Boltzmann provided a theoretical derivation of the SB law on the basis of classical physics (i.e. Maxwell electrodynamic equations and first and second principles of thermodynamics), without introducing any quantum argument. 
On the contrary, the well known attempt made by Rayleigh and Jeans to obtain the SB law from classical-physics  leads to the ultraviolet catastrophe.
We remark that,
the  Boltzmann derivation of the SB law is reported today in the literature without any proper comment on the above paradox, apart from a recent paper by Montambaux in 2018 \cite{montambaux18}.
\par
The aim of this paper is to retrace the Boltzmann theoretical derivation of the Stefan law by shedding light on the hidden assumptions concerning the quantum mechanical and relativistic principles related to the quantization of the energy spectrum (Planck 1901 \cite{planck01}), the development of ensemble statistics \cite{gibbs02}, the concept of photons coined by Einstein in 1905, the  existence of a zero-point energy (Planck 1912 \cite{planck12} and Casimir 1948 \cite{casimir48}). 
Remarkably, the quantum energy spectrum described in terms of a photon gas, as responsible of the  Wien displacement law (1893), allows for the vanishing of an ultra-violet catastrophe, while the Casimir effect, associated with the zero-point energy, allows  for the vanishing of a vacuum  catastrophe \cite{adler95}. 
In so doing, we remind that, the SB law is associated with the $T^3$ increase of the average number of photons that are present in the black-body cavity at a given temperature, a quantum effect due to the Boson nature of photon statistics. 
In this way, the Boltzmann paradox is solved by noticing the explicit assumption Boltzmann made of an internal energy for the physical system depending from both temperature and volume that implicitly leads to the non conservation of the number of electromagnetic 
(EM) modes for an isothermal transformation \cite{montambaux18}. 
Both the assumptions will be justified within a quantum mechanical approach.
In other words, the Boltzmann derivation of the SB law cannot be claimed to be obtained within a classical approach but from ad hoc visionary conjectures that accounted for the experimental evidence suggested by Stefan \cite{montambaux18}.
The present results complement recent papers that revisited the fluctuation dissipation theorem published by the same Authors \cite{reggiani18,reggiani20}.
\section{Theory}
As physical system we consider a box  with cross section $A$ and length $L$,  with one piston 
free to move thus enabling the control of the pressure of the radiation inside the box and its volume $V=A L$.
Following Planck work, we assume that the box behaves as an ideal black-body cavity with the walls, made by ideal conducting metals at fixed temperature $T$, acting as thermal reservoir with an internal vacuum obtained by pumping out all the material inside (for example gases, and residues of materials different from the walls) \cite{robitaille15}.  In the Boltzmann classical approach, the vacuum is thought to be  filled by a set of EM plane waves, linearly polarized, that do not interact each others,  emitted and absorbed by the walls at the given temperature under thermal equilibrium conditions. We notice that because of the random process of emission and absorption from the walls, the number of plane waves, $N$, is not fixed. As a consequence, from a thermodynamic point of view, the chosen physical system acts as a grand-canonical ensemble.
\par
The following Sects. 2.1 and 2.2   briefly outline the classical and quantum derivations of the SB law according to a modern approach of classical and quantum statistics.  
To this purpose, the classical approach basically follows the 
original  Boltzmann derivation, see Ref. \cite{boltzmann},  including  concepts  hidden, or not yet known, by  Boltzmann, such as: the assumption that the EM fluid internal energy takes a finite value and is a function of temperature and volume, the use of a grand-canonical ensemble, the chemical potential, the Poynting vector, etc.  
By contrast, the quantum approach includes the Planck distribution (necessary to provide the microscopic definition of the internal energy and associated properties of the photon gas model)  and the Casimir effect associated with the zero-point energy contribution.
\subsection{Classical electro-magnetic model}
Following Maxwell equations, each EM wave (or mode) is  characterized by a given frequency, 
$f$, and a wavelength, $\lambda$, satisfying $c=f \lambda$.
By using the Poynting vector, $\vec{S}$,
\begin{equation}
\vec{S} = \frac{\vec E}{\mu_0} \times \vec B
\label{eq4}
\end{equation} 
with $\vec E$ and $\vec B$, respectively  the electric and magnetic fields perpendicular each other, laying in a plane perpendicular to the direction of the wave individuated  by the direction of the Poynting vector, and  $\mu_0$ the vacuum magnetic permeability.
\par 
The  energy dispersion of the mode writes:
\begin{equation}
\epsilon=cp
\label{eq5}
\end{equation}
with $\epsilon$ the mode energy,  $p$ the modulus of the mode momentum, and
the average energy, taken on a period of the wave, $U^c(E_0,B_0)$ :
\begin{equation}
U^c(E_0,B_0)
=\frac{1}{2} \int_V (\epsilon_0 E_0^2 + \frac{B_0^2}{\mu_0}) dV
\label{eq6}
\end{equation}
where $\epsilon_0$ is  the vacuum dielectric permittivity, $E_0$ and $B_0$ are the maximum amplitude of the electric and  the magnetic field, respectively, and the integral is extended over the volume of the physical system.
We notice that the average energy of the classical single mode expressed in Eq. 6 is compatible with the assumption of an average energy dependent on volume and temperature.
Indeed, by recalling that, following energy equipartition both the quantities $E_0^2$ and $B_0^2$ can be expressed in terms of $T$, it is $U^c(E_0,B_0) = U^c(T,V)$.
\par
For an ensemble of EM modes uniformly distributed inside the 
physical system the state equation for this EM fluid is \cite{bartoli1884}:
\begin{equation}
P^c = \frac{U^c_{Tot}}{3V}
\label{eq7}
\end{equation}
with $P^c$ the pressure exerted by the ensemble of modes on the walls of the physical system  and 
\begin{equation}
U^c_{Tot} = \sum U^c
\label{eq8}
\end{equation}
the total internal energy of the ensemble of modes with the sum extended over all the expected number of modes, $N$, in the physical system. We notice that in agreement with Rayleigh-Jeans $N$, and so $U^c_{Tot}$,  would diverge.
\par
By writing the  first and second law of thermodynamics in differential form for a  grand canonical ensemble:
\begin{equation}
\frac{dU^c_{Tot}}{dV}
 = T\frac{dS^c}{dV} - P^c  + \mu\frac{dN}{dV} 
\label{eq9}
\end{equation}
where $S^c$ is the entropy of the ensemble of EM modes and $\mu$ the corresponding chemical potential.
\par
By recalling that $dS^c/dV=dP^c/dT$ and assuming with Boltzmann $\mu dN=0$ and that 
$U^c_{Tot}=U^c_{Tot} (V,T)$, that is, the total internal energy  is function of volume and temperature, it is obtained:
\begin{equation}
\frac{4 P^c}{dP^c} = \frac{T}{dT}
\label{eq10}
\end{equation}
The solution of the above equation gives
\begin{equation}
P^c= aT^4
\label{eq11}
\end{equation}
\begin{equation}
U^c_{Tot} =3 a V T^4
\label{eq12}
\end{equation}
with $a$ an integration constant determined by experimental data and/or by the appropriate statistical laws for the ensemble of EM modes.
\par
Equation (\ref{eq12}) written in terms of the internal energy density $u=dU^c_{Tot}/dV$ is known as the SB law with the constant $12a/c = \sigma$ being
the SB fundamental constant that is here obtained from the
physics known up to the 19th century.
\par
The drawbacks of the above Boltzmann derivation are:

i) the theory does not give the properties of the EM fluids necessary to provide a physical interpretation of the SB constant,

ii) the attempts to provide a statistical model for the number of modes failed by producing a divergence of the radiant emittance, the so called ultraviolet catastrophe,

iii) the impossibility to interpret the shape of the black-body radiation spectrum available from experiments,

iv) the difficulty to justify the two implicit Boltzmann assumptions: 1)  $\mu dN=0$ and, 
2) $U^c_{Tot}=U^c_{Tot} (V,T)$.

We remark that assumption iv (1) might be understood since at the time of  Boltzmann formulation the grand canonical ensemble, and thus the concept of a chemical potential, where not yet introduced, in any case the condition of a fixed number of particles, say $dN=0$,  for the thermodynamics of a classical gas, leading also to the condition $\mu dN=0$, was a standard one.
The assumption iv (2) is more subtle. Indeed, for a classical massive gas the internal energy was known to depend only from the temperature, and thus $dU^c_{Tot}/dV = u=0$ by definition. However, for the case of a gas of EM modes there was no physical reason to justify that the internal energy is function of both thermodynamic variables, temperature and volume, which is essential to reproduce the SB law.  As will be shown in the following, only quantum statistics will prove such an explicit dependence of the internal energy of a gas of photons by providing its explicit expression.
Therefore,
stating that the SB law is a result of classical physics is at least not correct.  As a counter proof, by using the same procedure for the classical massive gas, where apart from a factor of 2 the similar relation in Eq. (\ref{eq7})  holds, Boltzmann procedure predicts an internal-energy increase with temperature as $T^{5/2}$  instead of the known increase as $T$ \cite{leff02}. 
\subsection{Quantum photon model}
The microscopic ingredients to obtain the SB law in terms of a quantum photon gas replacing the classical EM model are: 
\par
(i) the quantization of the electromagnetic energy for a single photon given by:
\begin{equation}
\epsilon = hf (n + \frac{1}{2})
\label{eq13}
\end{equation}
with $n=0, 1, 2, ...$ a quantum number and, 
\par
(ii)
the definition of the average energy for a single photon $U^q(f,T)$ as:
$$
U^q(f,T) = U^q_{Planck}(f,T) + U^q_{ZP}(f)
$$
\begin{equation}
=\frac{hf}{exp(hf/K_BT)- 1} + \frac{hf}{2} 
\label{eq14} 
\end{equation}
\par
We stress that the split into two contributions of $U^q(f,T)$ as given  in the r.h.s. of Eq. (\ref{eq14}) is of most physical importance, since each contribution represents a channel giving macroscopic and exclusive evidence of the average energy per photon associated with quantum electrodynamics.
Indeed, the first-Planck contribution represents a property of the coupling between the thermal reservoir and the photon gas at equilibrium under thermodynamic limiting conditions.
As such, this contribution is a universal function of the temperature which takes a finite value at any frequency.
Accordingly, it vanishes at $T=0$,  it is independent of the external shape of the physical system and by definition it involves only the photons inside the physical system. Its spectrum can be directly measured by standard experimental techniques in a wide range of frequencies, typically from mHz to THz, and excellent agreement between theory and experiments is a standard achievement.
By contrast, the second zero-point (ZP) contribution is  associated with the zero-point energy, it is a quantum property of the vacuum, and by definition it involves all the photons in vacuum, i.e. inside and outside the physical system.  
Accordingly, its value diverges with frequency, it does not vanish at
$T=0$,  it has never been measured directly  but only through its effects, and in particular the Casimir effect that evidences  an attractive or repulsive force acting between finite parts of the physical system \cite{gong20}, as predicted by Casimir in 1948 \cite{casimir48} for the simple case of two parallel perfect conducting plates.
(For an updated bibliography on the Casimir effect the reader can refer to
J. Babb, bibliography on the Casimir Effect web site, 
https://www.cfa.harvard.edu/~babb/casimir-bib.html)
\par
The essential difference between the above two contributions is better explained when considering the total energy $U^q_{Tot, ZP}$ obtained by summation over all the photon modes of each contribution in the presence of a physical system.
For the Planck term, the summation is easily performed and gives the well-known Planck law of 1901 that leads to the SB law for the total radiation intensity emitted by an ideal black-body cavity at a given temperature:
\begin{equation}
U^q_{Tot,Planck} = \sum K_BT \ \frac{x}{e^x-1} 
= 3 \frac{\zeta(4)}{\zeta(3)} N K_B T
\label{eq15} 
\end{equation}
where: the sum is extended over all the photon modes inside the physical system,  $\zeta(n)$ is the Riemann $\zeta$ function, and
$\Gamma(3) \zeta(3) =\int_0^{\infty} x^2/(e^x-1) dx = 2.404$.
\begin{equation}
N(V,T)=8 \pi\Gamma(3)\zeta(3) V (\frac{K_B T}{ch})^3 
=2.03 \times 10^7 \ V T^3
\label{eq16}
\end{equation}
is the average number of photons inside the volume of the physical system  expressed in $m^3$. Notice the dependence of the photon number from the volume and the temperature.
As a consequence, it is
\begin{equation}
U^q_{Tot,Planck}=\frac{4V}{c}q = \frac{4}{c} \sigma V T^4
\label{eq17} 
\end{equation}
We notice that the equation above coincides with the SB law, showing the explicit temperature and volume dependence of the photons internal-energy as tacitly assumed by Boltzmann in his derivation of the SB law.
\par
For the zero-point term the  summation gives the expectation value of the total energy of the EM field in vacuum as: 
\begin{equation}
U^q_{Tot,ZP} = \frac{1}{2} \sum hf \rightarrow \infty
\label{eq18} 
\end{equation}
The sum gives a divergent energy contribution  when evaluated all over the space, which leads to the so-called vacuum catastrophe \cite{adler95}. 
Otherwise, as observed by Casimir \cite{casimir48}, real measurements are
performed on finite-size systems where manifestations of zero-point energy are directly observable \cite{kardar99}. 
In particular, the different content of EM energy
inside and outside an assigned region produces a finite value of the zero-point total energy. 
In general, Eq. (\ref{eq18}) is solved by using specific boundary conditions 
related to: 
(i) the shape of the physical system, and
(ii) the material used to construct the physical system.
\par
Calculations of the Casimir force are in general not easy to be performed \cite{ederly06,schmidt08,auletta09}, and here we report the simple but significant  case considered by Casimir \cite{casimir48} and further confirmed by more detailed mathematical approaches \cite{milton} of two opposite conducting plates with surface $A$ and distance $L$: 
\begin{equation}
U^q_{ZP,Casimir} 
=- \frac{\pi h c A}{1440 L^3} 
\label{eq19} 
\end{equation}
\par
The negative value of the Casimir energy implies an attractive force, the so called Casimir force, $F_{C}$, acting between opposite conducting plates, given by
\begin{equation}
F_{C}
= - \frac{\partial U^q_{ZP,Casimir}}{\partial L}
=- \frac{\pi h c A}  {480 L^4} 
\label{eq20} 
\end{equation}
The Casimir prediction of 1948 has been successively validated experimentally with increasing degree of accuracy. 
As a consequence of this quantum force, the physical system is no longer mechanically stable and the two opposite conducting plates 
might tend to implode when left free to move \cite{lebowitz69}.
\par 
For the present purposes, it is convenient to introduce the Casimir pressure, $P^q_C$, defined as
\begin{equation}
P^q_C=\frac{F_C}{A} 
= -\frac{\pi  h c}{480}L^{-4}
=-13.0 \times (10^{-7} \  L)^{-4} \ Pa
\label{eq21} 
\end{equation}
with $L$ in $m$.
The minus sign indicates that the Casimir pressure is exerted from the outside to the inside of the given metallic plates when left free to move one with respect to the other, thus acting on a piston that can alter the volume of the container in a quasi static way. 
We remark that, being $P^q_C = - 13.0 \ Pa$  for $L = 0.1 \ \mu m$, the Casimir pressure   becomes of some significance for distances below about the micrometer size where deviations of real system from the ideal black-body model are expected and confirmed \cite{decca11}. 
\par
We can go further in revisiting the SB law by noticing the presence of relativistic and quantum effects in terms of the existence of a radiation pressure, $P^q_{R}$, exerted by the photon gas on  the internal walls of the physical system, which algebraically adds to the Casimir pressure exerted by the zero-point energy as defined in the previous section.
The radiation pressure, experimentally measured in Refs. \cite{lebedev1901,nichols1903}, is  
given (in Pascal units)
by:
\begin{equation}
P^q_{R}= P^c_{R}=\frac{4  \sigma}{3 c}T^4
= \frac{8 \pi^5 K_B^4}{45 h^3 c^3} T^4
=2.52 \times 10^{-16} \ T^4  \ Pa
\label{eq23} 
\end{equation}
We remark that $P^q_R = 2.52 \ Pa$ for a temperature $T= 10^4 \ K$, a value higher for about a factor of three  to implement a realistic black-body cavity, that in any case tends to balance the Casimir pressure.
We conclude that both the Casimir and radiation pressure can interfere in the region of lengths and temperatures considered above. 
In general, the total pressure exerted on the piston of the physical system due to the photon gas, $P^q_{Tot}$, is  given by
\begin{equation}
P^q_{Tot}= P^q_R + P^q_C
\label{eq25} 
\end{equation}
For an ideal black-body cavity the first radiation-contribution is universal and function of the temperature only, as predicted by the SB law and confirmed by the Planck law. The second Casimir contribution is associated with the zero-point energy and  depends on the material forming
the physical system and its geometry. Here we report the value estimated by Casimir model that has been validated by several experiments.  However, further investigations should better detail results for different materials and geometry.
To this purpose, further experiments for the determination of the radiant emittance at sufficiently high temperatures, let us say  above about $1000 \ K$  and/or  small size physical system, let us say below about $ 1 \ \mu m$ length, should confirm some of the expectations on the Casimir pressure given in this paper.
\section{Conclusion and remarks}
In the present paper we propose that, in a modern  derivation of the  SB law, the 
quantum-relativistic nature of the photon gas should be explicitly considered.
To this purpose,  for the physical system the measurement of the pressure exerted by the photon gas on the walls as function of the temperature and the distance between two parallel plates of different materials could provide useful information on 
the Casimir force associated with different materials.
\par
In most cases (e.g. temperature below about $1000 \ K$ and  macroscopic systems over about   $1\mu m$ size), both the Casimir and  the radiation pressures are negligible with respect to  the elastic limits of a rigid physical system.
Therefore,  these forces can be easily absorbed by the elastic properties of the environment in which the physical system  is embedded.
However, for microscopic systems (i.e. in the case of length scales below about $ 1 \ \mu m$) and/or sufficiently high temperatures above about $1000 \ K$, Casimir as well as radiation pressures  should become relevant enough to lead
to significant deviations from the SB law, and/or Planck emission spectra, as already 
evidenced by experiments in Refs. \cite{reiser13,sister17}.
In short, a modern physical derivation of the SB law should imply an explicit justification of two basic assumptions that are lacking in the original Boltzmann derivation: 

(i) a zero value of the chemical potential associated with the photon-gas model of the electromagnetic field, in other words the average number of photons as given by Planck law in Eq. (\ref{eq15}), should be considered as a function  of volume and temperature defining the state equation of the photon gas together with Eqs. (\ref{eq11},\ref{eq12});

(ii) the presence of a pressure originated by the sum of two contributions belonging to  the radiation and the Casimir effects.
\par
Assumption (i) follows from the property that under thermal equilibrium conditions the number of photons in the physical system is not fixed.
Assumption (ii) follows from the quantum effects associated with the zero-point energy implying the existence of a Casimir force between different parts of the physical system.
Therefore, the claiming that SB law is obtained on the basis of classical physics is not correct since the  Boltzmann derivation implicitly includes thermodynamic properties justified only in terms of quantum mechanical models as discussed in the present paper.
\par
Even if outside the scope of this paper, we would remark that by relaxing the ideal black-body cavity and Casimir models used here, deviations from a universal behavior of the SB as well as of Planck and Casimir laws to account for size, temperature, and material characteristics have been evidenced and theoretically analyzed, see for example \cite{decca11,thompson18}.  
\end{document}